\documentclass[prb,twocolumn,showpacs,amsmath,amssymb,superscriptaddress,floatfix]{revtex4}

\usepackage[english]{babel}
\usepackage{amsfonts}
\usepackage{graphicx}

\begin{document}

%\title{Electron-phonon coupling lowers bipolaron kinetic energy in the t-J Holstein model}
\title{Gain of the kinetic energy of bipolarons in the $t$-$J$-Holstein model based on electron-phonon coupling}

\author{L. Vidmar}
\affiliation{J. Stefan Institute, 1000 Ljubljana, Slovenia}

\author{J. \surname{Bon\v ca}}
\affiliation{Faculty of Mathematics and Physics, University of Ljubljana, 1000
Ljubljana, Slovenia}
\affiliation{J. Stefan Institute, 1000 Ljubljana, Slovenia}

%\author{S. \surname{Maekawa}}
%\affiliation{Institute for Materials Research, Tohoku University,
%Sendai 980-8577, Japan}
%\affiliation{CREST, Japan Science and
%Technology Agency (JST), Tokyo 102-0075, Japan}

%\author{T. \surname{Tohyama}}
%\affiliation{Yukawa Institute for Theoretical Physics, Kyoto
%University, Kyoto 606-8502, Japan}

%\author{P. \surname{Prelov\v sek}}
%\affiliation{Faculty of Mathematics and Physics, University of
%Ljubljana, Ljubljana, Slovenia} \affiliation{J. Stefan Institute,
%Ljubljana, Slovenia}

\date{\today}
\begin{abstract}
With increasing  electron-phonon  coupling as described within the  $t$-$J$-Holstein model, bipolaron kinetic energy   is lowered in comparison with that of the polaron. This effect is accompanied with "undressing" of bipolaron from lattice degrees of freedom. Consequently, the effective bipolaron mass becomes smaller than  the   polaron mass.  Magnetic as well as lattice degrees of freedom cooperatively contribute to formation of  spin-lattice bipolarons. 
\end{abstract}

%  Bipolarons  nonconventional mechanisms in superconductivity   74.20.Mn   ***
%  Bipolarons electronic structure of solids                                       71.38.Mx    
%  Electron-phonon interactions electronic structure of solids,          71.38.-k     ***
%  Pairing symmetries (superconductivity),                                       74.20.Rp
%  t-J model,                                                                                      74.20.-z     ***
%

\pacs{74.20.Mn, 74.20.-z, 71.38.-k} \maketitle

\section{Introduction}

In strongly correlated systems superconductivity (SC) may occur as a consequence of the kinetic energy lowering as opposed to a standard BCS-type superconductor  where there is a slight kinetic energy raise accompanied by the  lowering of the potential energy that stabilizes  the  SC state \cite{hirsch,hirsch162,wheatley}. Experimentally, kinetic energy lowering  would reflect in the violation of the low-frequency optical sum rule and lead to a change of high-frequency optical absorption upon entering SC state \cite{hirsch201,molegraaf}.

The idea of in-plain kinetic energy driven pairing, originally  proposed by J.E. Hirsch,   is based on the Hubbard-type  model where hopping depends upon the occupation number \cite{hirsch162}. The possibility of kinetic energy gain has been investigated as well in more general models with correlated electrons, such as the standard Hubbard model.  Using variational Monte Carlo method authors of Ref.\cite{yokoyama} find kinetic energy gain in SC state with a $d$-wave symmetry  above a critical value of $U_c$. More recent calculations based on the dynamical cluster approximation show that pairing is driven by the kinetic energy gain \cite{jarrell_92}.

The main motive for the kinetic energy driven pairing in models possessing at least short range antiferromagnetic correlations relies on the argument that the motion of the single hole is obstructed  due to formation of strings of misaligned spins left in the wake of the propagating hole. A pair  of holes that propagates coherently should lower its kinetic energy as one hole moves in the wake created by the other hole.  This naive argument was challenged   by S.A. Trugman \cite{trugman} who suggested,  that  a pair of holes is less mobile than originally anticipated when assuming  the simple  string argument. Lower pair mobility occurs   due to  a frustration effect, which arises  from the fermion exchange processes. Cluster dynamical mean field studies \cite{haule}  on the $t$-$J$ model nevertheless show a small kinetic energy gain in the underdoped regime as the system enters  SC state, while there is a slight kinetic energy raise  in the overdoped regime. Authors of Ref.~\cite{wrobel} have demonstrated  the existence of the kinetic energy driven superconductivity in the $t$-$J$ model using the spin polaron technique. 

A scenario of the kinetic energy gain upon pair formation  is inherently connected with increased pair mobility  and consequently with lowering of the effective mass. In Holstein-type models, however,  the effective bipolaron mass is in the strong electron-phonon (EP) coupling regime  typically much  larger than the polaron one. A heavy bipolaron effective mass represents one of the main obstacles for bipolaronic theory of superconductivity \cite{alexandrov}. Addition of  Coulomb interaction \cite{bonca5} and generalization to physically more relevant Fr\" ohlich type EP interaction \cite{alexandrov2,bonca6, hague} contribute  to a substantial decrease of the bipolaron mass. Nevertheless, the bipolaron  remains heavier  in comparison with the polaron. 

While it is widely accepted that strong correlations govern  the physics of   high-$T_c$ superconductors \cite{dagotto_rev}, the notion of the importance of lattice effects with the emphasis on their role in formation of the SC state  is as well gaining momentum \cite{gunnarsson}.  
In this letter  we compare physical properties of systems with one  and two holes coupled to quantum phonons doped in the Heisenberg antifferomagnet. 
%As two polarons form a bound state
%We will mostly focus on the question wether there is  a lowering of the kinetic energy as two polarons %form a bound  a bipolaron state. %More specifically, 

We first show that there is no kinetic energy gain upon bipolaron formation in the pure $t$-$J$ model. Throughout   this paper we use the term kinetic energy for the expectation value of the hopping term in the $t$-$J$ model. We should point out that this term represents only the kinetic energy of the lower Hubbard band, as already noted by Wr\'obel {\it et al.}  in Ref.~\cite{wrobel}. Switching on EP coupling we  discover that the interplay between the kinetic, magnetic, and elastic energy leads to a formation of a spin-lattice bipolaron that in the crossover regime between the weak and strong EP coupling gains the kinetic energy relative to its polaron constituents. In this regime the effective mass of the spin-lattice bipolaron is decreased   relative to the effective mass  of the polaron. 

\section{Model and Method}

We employ  a recently developed method to solve  system of one  and two holes in the $t$-$J$ model defined on an infinite two-dimensional lattice, coupled to lattice degrees of freedom \cite{bonca4,bonca3,bonca2}.  We  investigate the influence of EP coupling via  the simplest extension of the $t$-$J$ model where holes  couple to dispersioneless  phonons. The model describes the influence of  apex oxygen vibration on doped holes, propagating in lightly doped CuO planes  in cuprates:
\begin{eqnarray}
H&=& -t\sum_{\langle {\bf i,j}\rangle,s}(\tilde c^\dagger_{{\bf i},s} \tilde c_{{\bf j},s} +\mathrm{H.c.}) +
J\sum_{\langle {\bf i,j}\rangle }( {\bf S}_{\bf i} {\bf S}_{\bf j} - \frac{1}{4}n_{\bf i} n_{\bf j} ) \nonumber \\
  &+& g \sum_{\bf i}n_{\bf i}^h(a_{\bf i}^+ + a_{\bf i})+\omega_0\sum_{\bf i}a_{\bf i}^+  a_{\bf i},\label{ham}
\end{eqnarray}
where $\tilde c_{{\bf i},s} = c_{{\bf i},s}(1 - n_{{\bf i},-s})$ is a projected  fermion operator, $t$ represents nearest neighbor overlap integral, the sum $\langle \bf i,j \rangle$  runs over pairs of nearest neighbors, $a_{\bf i}$  are phonon annihilation operators and $n_{\bf i} = \sum_s n_{{\bf i},s}$. $g$ and $\omega_0$ represent EP coupling constant and the Einstein phonon frequency, respectively.

While the numerical method has been in detail described in previous works \cite{bonca4,bonca3,bonca2}, we only briefly highlight  a few most relevant elements of the method. 
The construction of the functional space   for one and two holes starts from a N\' eel state with one or   two holes located on neighboring Cu sites and with zero phonon quanta.  In the case of a high symmetry point at ${\bf k} = (0, 0)$, the parent state of two holes can be chosen to exhibit a point-symmetry, belonging to a particular irreducible representation of the point group $C_{4v}$. Such a state is expressed as 
%
%\begin{equation}
$%
\vert \phi^{(0,0)}{\rangle}_a = \sum_{\boldsymbol{\gamma}}(-1)^{M_a(\boldsymbol{\gamma})}
c_0c_{\boldsymbol{\gamma}}\vert {\rm Neel };0\rangle,
$
%\label{parent}
%\end{equation}
%
where sum runs over four nearest neighbors in the case of $d-$ and $s-$ wave symmetry and over two in the case of $p_{x(y)}$-wave while $M_a (\boldsymbol{\gamma})$, $a\in \{ d,s,p \}$ sets the appropriate sign.

We generate new parent states by applying the generator of states 
$
\left \{ \vert \phi_l^{(N_h,M)}{\rangle}_a\right \} = \left ( H_{\rm kin} + 
H_{g}^M\right )^{N_h} \vert \phi^{(0,0)}{\rangle}_a
$ where
$H_{\rm kin}$  represent the first term in Eq.~\ref{ham}, $H_{g}$  represents 
third  term in Eq.~\ref{ham}. When parameter $M>1$ is chosen, the functional generator creates states with additional phonon quanta. This approach   ensures good convergence in the strong EP coupling regime where the ground state contains multiple phonon excitations \cite{bonca3}. In most cases we have used
$N_h = 6$ and $M=8$ that lead to $N_{\rm st} = 24 \times  10^6$ states. Full Hamiltonian
in Eq.~\ref{ham} is diagonalized within this limited functional space taking explicitly into account translational symmetry. As of now  we refer to one and two hole states  as polaron and bipolaron, where  polaron (bipolaron) signifies a hole (two holes), dressed with spin as well as lattice excitations. 

\begin{figure}[htb]
\includegraphics[width=11cm,clip]{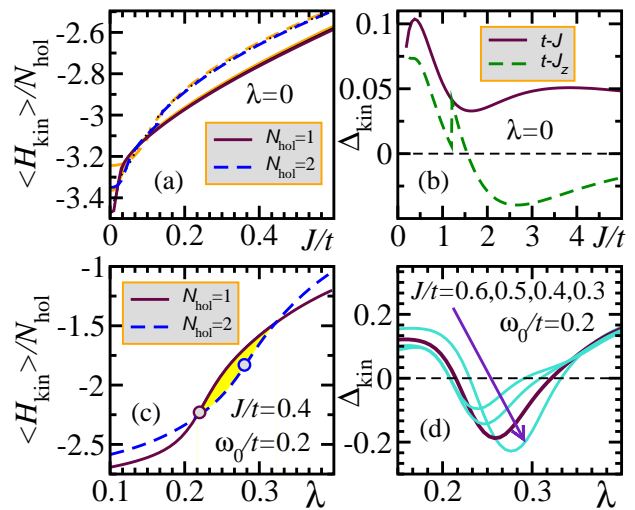}
\caption{(Color online) %
(a) $\langle H_{\rm kin}\rangle/N_{\rm hol}$ for the $N_{\rm{hol}}=1$ (polaron) and 2 hole (bipolaron)  in the $t$-$J$ model with $\lambda=0$. There are three nearly overlapping curves for each system obtained using the following parameters $N_h=10, 12$, and 14 with $M=0$ in all cases that lead to the following numbers of basis states: $N_{\rm st}=3.1\times 10^5, 2.6\times 10^6$, and $21 \times 10^6$ for the $N_{\rm hol}=2$ case.   Numbers of polaron states ($N_{\rm hol}=1$) are  typically a factor of 4 smaller; %Insert in a) shows  comparison of $\langle H_{\rm kin}\rangle/N_{\rm hol}$ for the $t$-$J_z$ model with $N_h=14, %M=0$;
(b)  $\Delta_{\mathrm{kin}}$ vs. $J/t$ for the $t$-$J$ model (full curve) and for the anisotropic $t$-$J_z$ model (dashed line). A discontinuity  in the latter case marks the crossover from the $p-$wave bipolaron ground state at small $J_z/t\lesssim 1.2$ to  $d-$wave at large $J_z/t$; (c) comparison of $\langle H_{\rm kin}\rangle/N_{\rm hol}$ between $N_{\rm{hol}}=1$ and 2 systems vs. $\lambda$. In this and all subsequent plots we used generator of states with $N_h=6$ and $M=8$ that led to a Hilbert space with $N_{\mathrm{nst}}=24\times 10^6$ states and a  a  maximal number of phonons $N_{\mathrm{ph}}=48$, (d) $\Delta_{\mathrm{kin}}$ representing the difference of kinetic energies between bipolaron and polaron  as defined in the text vs. $\lambda$ for different values of the exchange interaction $J/t$. The thicker curve is for  $J/t=0.4$. 
}\label{fig1}
\end{figure}

\section{Results}

\subsection{$t$-$J$ model}

We first investigate the possibility whether within the framework  of the pure $t$-$J$ model a bipolaron  gains  the kinetic energy in comparison with a polaron.  To this effect we show in Fig.~\ref{fig1}(a)   the expectation value of the kinetic energy per hole, {\it i.e.} $\langle H_{\rm kin}\rangle/N_{\rm hol}$, vs. $J/t$  of the one and two-hole Hilbert space.  Except in the unphysically small $J/t\lesssim 0.15$
%, where results are due to  the proximity of the Nagaoka state inconclusive, 
we find no kinetic energy gain of the bipolaron state.  This result is in agreement with Trugman's suggestion \cite{trugman}.  The frustration that arises  from the fermion exchange processes impedes  bipolaron   motion. There is no such effect in the polaron case.  

A higher mobility of the bipolaron should be more pronounced in the case of the $t$-$J_z$ model due to a lack of spin-flip processes that erase pairs of overturned spins and contribute to the gain of the kinetic energy of polarons  in the isotropic case. For this reason we have investigated the difference between kinetic energies per hole between bipolaron and polaron $\Delta_{\mathrm{kin}}= \langle H_{\rm kin}\rangle^{(2)}/2- \langle H_{\rm kin}\rangle^{(1)}$ in a wider, even though unphysical range of $J/t$, see Fig.~\ref{fig1}(b).  We find $\Delta_{\mathrm{kin}}<0$ for $J_z/t\gtrsim 1.5$. The discontinuity  is a consequence of a crossing between $p-$wave symmetry of the pair at small $J_z/t$ to $d-$wave symmetry for $J_z/t\gtrsim 1.2$. In contrast,  in the isotropic $t$-$J$ model $\Delta_{\mathrm{kin}}>0$  in the whole expanded  $J/t$ regime, presented in Fig.~\ref{fig1}(b). 

On the more technical side we report on a test of the  convergence of our method. In Fig.~\ref{fig1}(a) we present nearly overlapping curves obtained using  three different Hilbert spaces, generated by $N_h=10,12,14$ and $M=0$. Note, that in our calculation the maximal allowed hole distance is: $l_{\mathrm max}=N_h+1=15$ in the case of the largest Hilbert space with no phonons. This  should be compared with exact diagonalization calculations on finite square  lattices with $N$ sites, where $l_{\mathrm max}=\sqrt{N/2}=4$ in the case of $N=32$ sites \cite{cherny1}.

\subsection{$t$-$J$-Holstein model}

We now switch on the EP coupling. In Fig.~\ref{fig1}(c) we plot the kinetic energy per hole vs. dimensionless EP coupling constant $\lambda = g^2/8\omega_0t$. At small $\lambda$ kinetic energies of polaron  ($N_{\mathrm{hole}}=1$) and bipolaron  ($N_{\mathrm{hole}}=2$)  increase linearly with  $\lambda$. Such behavior is characteristic for the weak EP coupling regime. In the regime  $0.22\lesssim \lambda\lesssim 0.32$ the kinetic energy (per hole) of bipolaron crosses below the kinetic energy of polaron. The onset of this regime coincides with the crossover to strong coupling regime of the spin-lattice polaron \cite{nagaosa,bonca3}. If  we roughly define the crossover to the strong EP coupling regime as a point of the steepest increase of  $\langle H_{\rm kin}\rangle$ vs. $\lambda$, we discover, that the polaron state enters strong EP coupling regime at smaller $\lambda_c^{(1)}\sim 0.22$ than the bipolaron state, where  $\lambda_c^{(2)}\sim 0.28$.  The two  inflection points  are as well indicated with open circles in Fig.~\ref{fig1}(c). 

The dependence of the kinetic energy gain on the magnetic exchange interaction $J/t$ is investigated in  Fig.~\ref{fig1}(d) where we follow    $\Delta_{\mathrm{kin}}$ vs. $\lambda$ using different values of the exchange interaction.  As $J/t$ increases the kinetic energy gain is reduced. This behavior is in contrast   with the anisotropic case where at $\lambda=0$, $\Delta_{\mathrm{kin}}$ becomes negative only at large $J_z/t$, see  Fig.~\ref{fig1}(b). The kinetic energy gain, observed in  the isotropic case,  is driven by the EP coupling.  It is positioned  in the crossover from weak to  strong EP coupling regime.  In terms of the magnetic exchange interaction  it is located   well within the physically relevant regime of $J/t\in [0.3,0.4]$.

\begin{figure}[htb]
\includegraphics[width=11cm,clip]{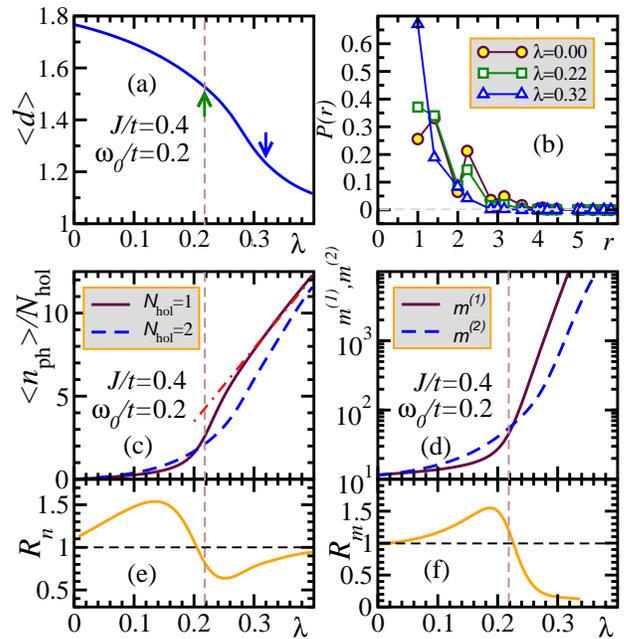}
\caption{(Color online) %
(a) Average hole distance vs. $ \langle d\rangle$.  Vertical dashed line indicates the onset of the regime of the kinetic energy gain, $\Delta_{\mathrm{kin}}< 0$ as seen in Fig.~\ref{fig1}(a).  (b) Probability $P(r)$ of finding holes at a distance of $r$ computed at three different values of $\lambda=0,0.22$ and  $0.32$, the latter two coinciding with positions of two arrows in (a), (c) $ \langle n_{\mathrm{ph}}\rangle/N_{\mathrm{hol}} $ vs. $\lambda$ of polaron  (full line) and bipolaron  system (dashed line). The dot-dashed line represents the fit of the polaron result to a straight line in the regime $\lambda>0.375$ that is given by $y=C + 8.8\lambda t/\omega_0$  (d) effective masses per hole $m^{(1)}$ and $m^{(2)}$ of polaron  (full line) and bipolaron (dashed line) system, (e) and (f)  $R_{\rm n}$ and $R_{\rm m}$ ratios  between phonon numbers and effective masses as presented in (c) and (d) respectively (see also deffinitions in the text). 
}\label{fig2}
\end{figure}

In Fig.~\ref{fig2}(a)  we present the  average hole distance 
$
\langle d\rangle = \sum_{r} r P(r),
$ 
of a bipolaron state,  where $P(r)$ represents the probability of finding a hole-pair at a distance of r: 
$
P(r) = \langle \sum_{\langle \bf i\not = j \rangle } n^h_{\bf i}n^h_{\bf j}
\delta\left [ \vert {\bf i-j}\vert -r\right ]  \rangle
/ \langle \sum_{\langle \bf i\not = j \rangle } n^h_{\bf i}n^h_{\bf j} \rangle.
$ 
At $J/t=0.4$ and $\lambda=0$ two holes form a bound bipolaron with the largest probability at a distance of $r=\sqrt{2}$, as consistent with previous calculations \cite{cherny1,eder,riera2}. With increasing EP coupling $\lambda$, $ \langle d\rangle $  experiences  the steepest decrease  in the middle of the regime where $\Delta_{\mathrm{kin}}<0$, {\it i.e.} for $0.22\lesssim \lambda\lesssim 0.32$. This is reflected in the change of the shape of  bipolaron where the hole distance with the largest probability $P(r)$ crosses over from $r=\sqrt 2$ at $\lambda=0$ to $r=1$ at $\lambda=0.32$, see Fig.~\ref{fig2}(b). It is somewhat counter-intuitive that such shrinking  of the bipolaron size simultaneously gives rise to  $\Delta_{\mathrm{kin}}<0$. Close proximity of holes namely restricts the range of hopping  that consequently leads to a raise of the kinetic energy, as as well seen in Fig.~\ref{fig1}(c). 

We gain additional insight into the mechanism leading to $\Delta_{\mathrm{kin}}<0$ by  presenting  comparison of the average number of phonons per hole between polaron and bipolaron. In Fig.~\ref{fig2}(c) we plot $ \langle n_{\mathrm{ph}}\rangle/N_{\mathrm{hol}} $ vs. $\lambda$. At small $\lambda$ the average $ \langle n_{\mathrm{ph}}\rangle/N_{\mathrm{hol}} $  is for bipolaron  slightly larger from that of the polaron,  while at $\lambda\sim 0.22$ it crosses below the polaron result.  In the large $\lambda$ limit both expectation values  should approach the strong coupling regime.   Indeed, $ \langle n_{\mathrm{ph}}\rangle/N_{\mathrm{hol}} $ for polaron for $\lambda \gtrsim 0.27$ approaches  a  straight line characteristic for the strong coupling result: $ \langle n_{\mathrm{ph}}\rangle=g^2/\omega_0^2 = 8\lambda t/\omega_0$, see the fit in Fig.~\ref{fig2}(c). The same is not true for the bipolaron, where $ \langle n_{\mathrm{ph}}\rangle/N_{\mathrm{hol}} $ is below  polaron result  for $\lambda\gtrsim 0.22$ and finally approaches the strong coupling result above $\lambda\gtrsim 0.35$. We should also note that in the $\Delta_{\mathrm{kin}}<0$ regime,  say around $\lambda=0.26$, $ \langle n_{\mathrm{ph}}\rangle/N_{\mathrm{hol}} $ for the polaron system exceeds result for bipolaron  by nearly  50\%. Results are consistent with the observation that the bipolaron crosses over to the  strong coupling regime within a wider crossover regime and at larger $\lambda$ than the polaron. 

In connection with  Fig.~\ref{fig2}(c) we note that in the vicinity of  the physically relevant values of $\lambda\in [0.22,0.32] $ the average number of phonons  in the bipolaron state is $ \langle n_{\mathrm{ph}}\rangle <20$. This is to be compared with the maximum number of phonon quanta contained in the Hilbert space $N_{\mathrm{ph}}=N_h*M=48$. We can conclude  that Hilbert space used in our method contains  sufficient amount of phonon degrees that empowers  our calculation reaching full convergence. 
To further investigate the difference  in the phonon number between bipolaron and polaron state we plot in  Fig.~\ref{fig2}(e) the corresponding ratio $R_{\rm n}= \langle n_{\mathrm{ph}}\rangle^{(2)}/2\langle n_{\mathrm{ph}}\rangle^{(1)}$ that shows 40\% decrease in the avergage phonon number per hole in the regime $\Delta_{\mathrm{kin}}<0$. Note that in the strong coupling regime $ R_{\rm n} \to 1$.
 
%at least in terms of generating  a sufficient number of phonon degrees of freedom. 

We proceed by presenting  comparison of effective polaron and bipolaron masses. The polaron dispersion of the $t$-$J$ model has a minimum at $\mathbf{k}=(\pi/2,\pi/2)$ and  is  highly anisotropic.  We  compute the  effective mass tensor in its  eigendirections $m_{\alpha \alpha}=2t\left (\partial^2E({\bf k})/\partial {\bf k} \partial {\bf k}\right )^{-1}_{\alpha \alpha}$. Taking into account anisotropic dispersion we define the polaron mass as $m^{(1)}=\sqrt{m_{||} m_\perp}$, were $m_{||}$ and $m_\perp$ are effective polaron masses along nodal and anti-nodal directions respectively. In contrast, bipolaron has  the energy minimum  at $k=0$ with  locally isotropic dispersion. We compute  effective bipolaron mass per hole from $m^{(2)}=m_{xx}/2$. In  Fig.~\ref{fig2}(d) we present $m^{(1)}$ and $m^{(2)}$ vs. $\lambda$. In the weak coupling regime we find the expected result where $m^{(2)}>m^{(1)}$. For $\lambda\gtrsim 0.23$ the opposite becomes true. Results in Fig.~\ref{fig2}(d) represent, at least to our knowledge, the first example where $m^{(2)}<m^{(1)}$. The ratio $R_{\rm m}=m^{(2)}/m^{(1)}$  in Fig.~\ref{fig2}(f) drops down to $R_{\rm m}\sim 0.2$ around  $\lambda\sim 0.3$.  This result is highly unusual. For comparison we draw attention to  a well known result for the  Holstein model where the effective mass of two particles forming a lattice bipolaron singlet, scales in the strong coupling regime as $m^{(2)}\propto \exp[4(g/\omega_0)^2]$ in comparison to $m^{(1)}\propto \exp[(g/\omega_0)^2]$ that leads to $m^{(2)}>>m^{(1)}$. In the Holstein Hubbard model, however, the on-site Coulomb interaction gives rise  to a formation of an   intersite bipolaron  with an effective mass that is comparable, nevertheless  always  larger than the polaron mass \cite{bonca5}. Recently Hague {\it et al.} \cite{hague} examined  a bipolaron defined on a triangular lattice and found unusually  small  effective bipolaron mass due to a crablike motion. Nevertheless, bipolaron effective mass remains  larger than that of the polaron. 

A note of caution: as seen in Fig.~\ref{fig2}(d) the absolute value of the effective bipolaron mass is rather large allready at the onset of the $\Delta_{\mathrm{kin}}<0$ regime,  $m^{(1)}\sim m^{(2)}\sim 60$ (in units of free electron mass) at $\lambda=0.23$. This shortcomming can be easily alleviated by taking into account physically more relevant longer-range Fr\" ohlich EP interaction that may reduce  $m^{(2)}$ and $m^{(1)}$up to an order of  magnitude \cite{alexandrov2,bonca6,hague}. Experimental results on the efective mass seem to be slightly ambiguous. Nevertheless, latest de Haas-van Alphen measurements of the cyclotron effective mass inside of the superconducting dome of YBCO show divergence with decreasing doping. They have measured effective masses as large as 4.5 free electron masses \cite{sebastian}. 

\begin{figure}[htb]
\includegraphics[width=11cm,clip]{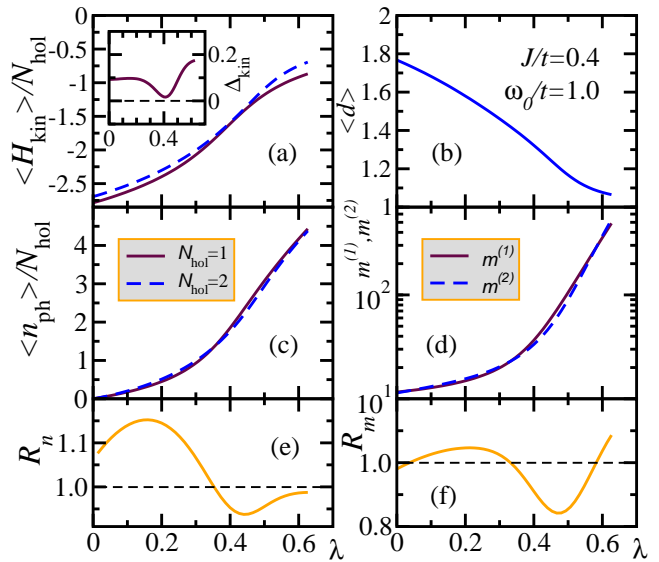}
\caption{(Color online) %
(a) $\langle H_{\rm kin}\rangle/N_{\rm hol}$ for $N_{\rm{hol}}=1$ and 2 systems vs. $\lambda$, 
(b) average hole distance $ \langle d\rangle$ vs. $\lambda$,    (c) $ \langle n_{\mathrm{ph}}\rangle/N_{\mathrm{hol}} $ vs. $\lambda$ of polaron  (full line) and bipolaron  system (dashed line),   (d) effective masses per hole $m^{(1)}$ and $m^{(2)}$ for polaron  (full line) and bipolaron (dashed line) system, (e) and (f)  $R_{\rm n}$ and $R_{\rm m}$ ratios  between phonon numbers and effective masses as presented in (c) and (d) respectively. Parameters of the model in all pictures (a, ..., f) are: $J/t=0.4$ and $\omega_0/t=1.0$.
}\label{fig3}
\end{figure}

In our search for deeper understanding of the phonon driven kinetic energy lowering and bipolaron mass renormalization we have investigated as well the regime with larger $\omega_0$, {\it i.e.} $\omega_0/t=1.0$.  In this case   the effect of EP coupling on the kinetic energy lowering disappears as seen from Fig.~\ref{fig3}(a) even though around $\lambda\sim 0.4$ $\Delta_{\mathrm kin}$ approaches zero, see also the insert of Fig.~\ref{fig3}(a). On the other hand  the effective bipolaron mass still shows a slight decrease with respect to the polaron mass in the regime  $0.35\lesssim \lambda\lesssim 0.6$,  as shown in Figs.~\ref{fig3}(d) and (e). This effect is closely connected to the average number of phonons  that shows barely detectable decrease in the same parameter regime, see  Figs.~\ref{fig3}(d) and (e).

\begin{figure}[htb]
\includegraphics[width=11cm,clip]{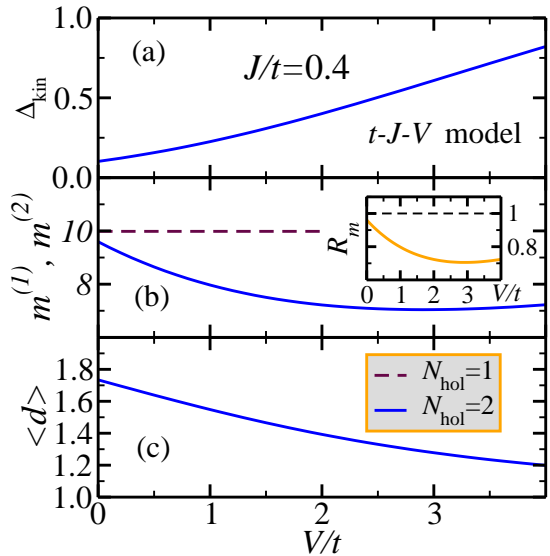}
\caption{(Color online) %
(a) $\Delta_{\rm kin}$  vs. $V/t$ for the $t$-$J$-$V$ model, (b) effective masses per hole $m^{(1)}$ and $m^{(2)}$ for polaron  (dashed line) and bipolaron (full line) system vs. $V/t$, and (c) average hole distance $ \langle d\rangle$ vs. $V/t$. We have used basis, generated by $N_{\rm h}=12$ and $M=0$. 
 }\label{fig4}
\end{figure}

\subsection{Attractive $t$-$J$-$V$ model} 

We have tried to reproduce the kinetic energy lovering and  the effective bipolaron mass renormalization using  a $t$-$J$-$V$ model where the effect of phonons is replaced by an effective  nearest neighbors attractive interaction $V$. To this effect we have added an attractive term of the form $H_V=-V\sum_{\langle {\bf i,j}\rangle}n^h_{\bf i}n^h_{\bf j}$ where $n^h_{\bf i}$ represents the hole-density operator, to a standard $t$-$J$ model. From Fig.~\ref{fig4}(a) we conclude that with increasing $V/t$ $\Delta_{\rm kin}$ remains positive and  monotonically  increases. We find  no kinetic energy gain in this simplified model. Nevertheless, we find a  decrease of the effective mass of  the bipolaron system as presented in Fig.~\ref{fig4}(b).  Effective mass ration  reaches  its minimum value  $R_{\mathrm m}\sim 0.75$ around   $V/t=3.0$. This should be compared to $R_{\mathrm m}\sim 0.13$  in the case of $\omega_0=0.2$ and $\lambda = 0.3$, Fig.~\ref{fig2}(f). Even though the effect of nearest neighbor attraction  $V$ on the average distance $\langle d \rangle$ is similar to the effect of increasing $\lambda$ (compare Figs~\ref{fig2}(a), \ref{fig3}(b), and \ref{fig4}(c)), the influence  of the EP coupling on hole motion in the $t$-$J$ model can not be entirely  explained using  an effective attraction between holes. Using the simplified model we nevertheless discover an important mechanism whereby the effective bipolaron mass decreases at fixed value of $J/t$ as the average distance between holes decreases by invreasing $V$.  In contrast to spin lattice bipolaron where its cumulative effective mass at $\omega_0=0.2$ and $\lambda\sim 0.3$ becomes lower than that of a polaron, $R_{\mathrm m}\sim 0.13<0.5$, spin bipolaron with attractive hole-hole interaction remains heavier than the polaron since $R_{\mathrm m}\sim 0.75 > 0.5$.

\section{Discussion}

The main mechanism behind the gain of the kinetic energy  as well as lowering of the effective mass of bipolaron emerges from the  competition  between kinetic, magnetic and lattice degrees of freedom. As two holes form a bipolaron   in the regime where $\Delta_{\mathrm{kin}}<0$, the system minimizes the kinetic energy  at the expense of increased  elastic energy. The gain of the kinetic energy contributes to energy splitting between states of  differnt point-group symmetries. This is in agreement with recent results of   Ref.\cite{bonca4} where  it has been shown, that EP coupling to transverse modes  in-plain oxygen vibrations stabilizes $d-$wave symmetry.  The increase of    the elastic energy emerges as  undressing of bipolaron from lattice degrees of freedom as seen as the decrease of $ \langle n_{\mathrm{ph}}\rangle/N_{\mathrm{hol}} $ below the respective polaron values. The gain in the kinetic energy does not represent the "glue" for the formation of the bipolaron since for $J/t\gtrsim 0.2$ bipolaron is already formed in the pure $t$-$J$ model  where no gain in the kinetic energy is found. Instead, it emerges as a side product of the competition between the kinetic, magnetic and elastic energies. 

Our results lead to a  novel paradigm where in a correlated system, coupled to quantum lattice degrees of freedom, upon pair formation  the bipolaron mobility increases due to a lower  effective mass as well as due to  a detectable gain in bipolaron kinetic energy. The attractive potential for binding of bipolaron appears as a  cooperative interplay  between magnetic and lattice degrees of freedom. 
While the original idea of the kinetic energy gain  as a mechanism for hole-pairing has been proposed for a system of correlated electrons, our finding opens the possibility where lowering of the kinetic energy in a correlated model is driven by the EP interaction.

\acknowledgments J.B. acknowledges stimulating discussions with I. Sega, C.D. Batista,  S.A. Trugman, and T. Tohyama  and the financial support of the SRA under grant P1-0044. 

%J.B. and L.V. acknowledge hospitality during their stay at IMR, Tohoku University, Sendai. 
%S.M. and T.T. acknowledge the financial support of
%the Next Generation Super Computing Project of Nanoscience
%Program, CREST, and Grant-in-Aid for Scientific Research (19052003) from MEXT. This work was %also upported by JPSJ and MHEST under the Japan-Slovenia Research Cooperative Program.
\bibliography{manukin}

\begin{thebibliography}{25}
\expandafter\ifx\csname natexlab\endcsname\relax\def\natexlab#1{#1}\fi
\expandafter\ifx\csname bibnamefont\endcsname\relax
  \def\bibnamefont#1{#1}\fi
\expandafter\ifx\csname bibfnamefont\endcsname\relax
  \def\bibfnamefont#1{#1}\fi
\expandafter\ifx\csname citenamefont\endcsname\relax
  \def\citenamefont#1{#1}\fi
\expandafter\ifx\csname url\endcsname\relax
  \def\url#1{\texttt{#1}}\fi
\expandafter\ifx\csname urlprefix\endcsname\relax\def\urlprefix{URL }\fi
\providecommand{\bibinfo}[2]{#2}
\providecommand{\eprint}[2][]{\url{#2}}

\bibitem[{\citenamefont{Hirsch}(2002)}]{hirsch}
\bibinfo{author}{\bibfnamefont{J.~E.} \bibnamefont{Hirsch}},
  \bibinfo{journal}{Science} \textbf{\bibinfo{volume}{295}},
  \bibinfo{pages}{2226} (\bibinfo{year}{2002}).

\bibitem[{\citenamefont{Hirsch and Marsiglio}(1989)}]{hirsch162}
\bibinfo{author}{\bibfnamefont{J.~E.} \bibnamefont{Hirsch}} \bibnamefont{and}
  \bibinfo{author}{\bibfnamefont{F.}~\bibnamefont{Marsiglio}},
  \bibinfo{journal}{Physica C} \textbf{\bibinfo{volume}{162}},
  \bibinfo{pages}{591} (\bibinfo{year}{1989}).

\bibitem[{\citenamefont{Wheatley et~al.}(1988)\citenamefont{Wheatley, Hsu, and
  Anderson}}]{wheatley}
\bibinfo{author}{\bibfnamefont{J.~M.} \bibnamefont{Wheatley}},
  \bibinfo{author}{\bibfnamefont{T.~C.} \bibnamefont{Hsu}}, \bibnamefont{and}
  \bibinfo{author}{\bibfnamefont{P.~W.} \bibnamefont{Anderson}},
  \bibinfo{journal}{Phys. Rev. B} \textbf{\bibinfo{volume}{37}},
  \bibinfo{pages}{5897} (\bibinfo{year}{1988}).

\bibitem[{\citenamefont{Hirsch}(1992)}]{hirsch201}
\bibinfo{author}{\bibfnamefont{J.~E.} \bibnamefont{Hirsch}},
  \bibinfo{journal}{Physica C} \textbf{\bibinfo{volume}{201}},
  \bibinfo{pages}{347} (\bibinfo{year}{1992}).

\bibitem[{\citenamefont{Molegraaf et~al.}(2002)\citenamefont{Molegraaf,
  Presura, van~der Marel, Kes, and Li}}]{molegraaf}
\bibinfo{author}{\bibfnamefont{H.~J.~A.} \bibnamefont{Molegraaf}},
  \bibinfo{author}{\bibfnamefont{C.}~\bibnamefont{Presura}},
  \bibinfo{author}{\bibfnamefont{D.}~\bibnamefont{van~der Marel}},
  \bibinfo{author}{\bibfnamefont{P.~H.} \bibnamefont{Kes}}, \bibnamefont{and}
  \bibinfo{author}{\bibfnamefont{M.}~\bibnamefont{Li}},
  \bibinfo{journal}{Science} \textbf{\bibinfo{volume}{295}},
  \bibinfo{pages}{2239} (\bibinfo{year}{2002}).

\bibitem[{\citenamefont{Yokoyama et~al.}(2004)\citenamefont{Yokoyama, Tanaka,
  Ogata, and Tsuchiura}}]{yokoyama}
\bibinfo{author}{\bibfnamefont{H.}~\bibnamefont{Yokoyama}},
  \bibinfo{author}{\bibfnamefont{Y.}~\bibnamefont{Tanaka}},
  \bibinfo{author}{\bibfnamefont{M.}~\bibnamefont{Ogata}}, \bibnamefont{and}
  \bibinfo{author}{\bibfnamefont{H.}~\bibnamefont{Tsuchiura}},
  \bibinfo{journal}{Journal of the Physical Society of Japan}
  \textbf{\bibinfo{volume}{73}}, \bibinfo{pages}{1119} (\bibinfo{year}{2004}).

\bibitem[{\citenamefont{Maier et~al.}(2004)\citenamefont{Maier, Jarrell,
  Macridin, and Slezak}}]{jarrell_92}
\bibinfo{author}{\bibfnamefont{T.~A.} \bibnamefont{Maier}},
  \bibinfo{author}{\bibfnamefont{M.}~\bibnamefont{Jarrell}},
  \bibinfo{author}{\bibfnamefont{A.}~\bibnamefont{Macridin}}, \bibnamefont{and}
  \bibinfo{author}{\bibfnamefont{C.}~\bibnamefont{Slezak}},
  \bibinfo{journal}{Phys. Rev. Lett.} \textbf{\bibinfo{volume}{92}},
  \bibinfo{pages}{027005} (\bibinfo{year}{2004}).

\bibitem[{\citenamefont{Trugman}(1988)}]{trugman}
\bibinfo{author}{\bibfnamefont{S.~A.} \bibnamefont{Trugman}},
  \bibinfo{journal}{Phys. Rev. B} \textbf{\bibinfo{volume}{37}},
  \bibinfo{pages}{1597} (\bibinfo{year}{1988}).

\bibitem[{\citenamefont{Haule and Kotliar}(2007)}]{haule}
\bibinfo{author}{\bibfnamefont{K.}~\bibnamefont{Haule}} \bibnamefont{and}
  \bibinfo{author}{\bibfnamefont{G.}~\bibnamefont{Kotliar}},
  \bibinfo{journal}{Europhys. Lett.} \textbf{\bibinfo{volume}{77}},
  \bibinfo{pages}{27007} (\bibinfo{year}{2007}).

\bibitem[{\citenamefont{Wr\'obel et~al.}(2003)\citenamefont{Wr\'obel, Eder, and
  Micnas}}]{wrobel}
\bibinfo{author}{\bibfnamefont{P.}~\bibnamefont{Wr\'obel}},
  \bibinfo{author}{\bibfnamefont{R.}~\bibnamefont{Eder}}, \bibnamefont{and}
  \bibinfo{author}{\bibfnamefont{R.}~\bibnamefont{Micnas}},
  \bibinfo{journal}{J. Phys.: Condens. Matter} \textbf{\bibinfo{volume}{15}},
  \bibinfo{pages}{2755} (\bibinfo{year}{2003}).

\bibitem[{\citenamefont{Alexandrov and Mott}(1994)}]{alexandrov}
\bibinfo{author}{\bibfnamefont{A.~S.} \bibnamefont{Alexandrov}}
  \bibnamefont{and} \bibinfo{author}{\bibfnamefont{N.~F.} \bibnamefont{Mott}},
  \bibinfo{journal}{Rep. Prog. Phys.} \textbf{\bibinfo{volume}{57}},
  \bibinfo{pages}{1197} (\bibinfo{year}{1994}).

\bibitem[{\citenamefont{Bon\v{c}a et~al.}(2000)\citenamefont{Bon\v{c}a,
  Katra\v{s}nik, and Trugman}}]{bonca5}
\bibinfo{author}{\bibfnamefont{J.}~\bibnamefont{Bon\v{c}a}},
  \bibinfo{author}{\bibfnamefont{T.}~\bibnamefont{Katra\v{s}nik}},
  \bibnamefont{and} \bibinfo{author}{\bibfnamefont{S.~A.}
  \bibnamefont{Trugman}}, \bibinfo{journal}{Phys. Rev. Lett.}
  \textbf{\bibinfo{volume}{84}}, \bibinfo{pages}{3153} (\bibinfo{year}{2000}).

\bibitem[{\citenamefont{Alexandrov and Kornilovitch}(1999)}]{alexandrov2}
\bibinfo{author}{\bibfnamefont{A.~S.} \bibnamefont{Alexandrov}}
  \bibnamefont{and} \bibinfo{author}{\bibfnamefont{P.~E.}
  \bibnamefont{Kornilovitch}}, \bibinfo{journal}{Phys. Rev. Lett.}
  \textbf{\bibinfo{volume}{82}}, \bibinfo{pages}{807} (\bibinfo{year}{1999}).

\bibitem[{\citenamefont{Hague et~al.}(2007)\citenamefont{Hague, Kornilovitch,
  Samson, and Alexandrov}}]{hague}
\bibinfo{author}{\bibfnamefont{J.~P.} \bibnamefont{Hague}},
  \bibinfo{author}{\bibfnamefont{P.~E.} \bibnamefont{Kornilovitch}},
  \bibinfo{author}{\bibfnamefont{J.~H.} \bibnamefont{Samson}},
  \bibnamefont{and} \bibinfo{author}{\bibfnamefont{A.~S.}
  \bibnamefont{Alexandrov}}, \bibinfo{journal}{Phys. Rev. Lett.}
  \textbf{\bibinfo{volume}{98}}, \bibinfo{pages}{037002}
  (\bibinfo{year}{2007}).

\bibitem[{\citenamefont{Bon\v{c}a and Trugman}(2001)}]{bonca6}
\bibinfo{author}{\bibfnamefont{J.}~\bibnamefont{Bon\v{c}a}} \bibnamefont{and}
  \bibinfo{author}{\bibfnamefont{S.~A.} \bibnamefont{Trugman}},
  \bibinfo{journal}{Phys. Rev. B} \textbf{\bibinfo{volume}{64}},
  \bibinfo{pages}{094507} (\bibinfo{year}{2001}).

\bibitem[{\citenamefont{Dagotto}(1994)}]{dagotto_rev}
\bibinfo{author}{\bibfnamefont{E.}~\bibnamefont{Dagotto}},
  \bibinfo{journal}{Rev. Mod. Phys.} \textbf{\bibinfo{volume}{66}},
  \bibinfo{pages}{763} (\bibinfo{year}{1994}).

\bibitem[{\citenamefont{Gunnarsson and Rosch}(2008)}]{gunnarsson}
\bibinfo{author}{\bibfnamefont{O.}~\bibnamefont{Gunnarsson}} \bibnamefont{and}
  \bibinfo{author}{\bibfnamefont{O.}~\bibnamefont{Rosch}},
  \bibinfo{journal}{Journal of Physics: Condensed Matter}
  \textbf{\bibinfo{volume}{20}}, \bibinfo{pages}{043201 (22pp)}
  (\bibinfo{year}{2008}).

\bibitem[{\citenamefont{Bon\v{c}a et~al.}(2008)\citenamefont{Bon\v{c}a,
  Maekawa, Tohyama, and Prelov\v{s}ek}}]{bonca3}
\bibinfo{author}{\bibfnamefont{J.}~\bibnamefont{Bon\v{c}a}},
  \bibinfo{author}{\bibfnamefont{S.}~\bibnamefont{Maekawa}},
  \bibinfo{author}{\bibfnamefont{T.}~\bibnamefont{Tohyama}}, \bibnamefont{and}
  \bibinfo{author}{\bibfnamefont{P.}~\bibnamefont{Prelov\v{s}ek}},
  \bibinfo{journal}{Phys. Rev. B} \textbf{\bibinfo{volume}{77}},
  \bibinfo{pages}{054519} (\bibinfo{year}{2008}).

\bibitem[{\citenamefont{Bon\v{c}a et~al.}(2007)\citenamefont{Bon\v{c}a,
  Maekawa, and Tohyama}}]{bonca2}
\bibinfo{author}{\bibfnamefont{J.}~\bibnamefont{Bon\v{c}a}},
  \bibinfo{author}{\bibfnamefont{S.}~\bibnamefont{Maekawa}}, \bibnamefont{and}
  \bibinfo{author}{\bibfnamefont{T.}~\bibnamefont{Tohyama}},
  \bibinfo{journal}{Phys. Rev. B} \textbf{\bibinfo{volume}{76}},
  \bibinfo{pages}{035121} (\bibinfo{year}{2007}).

\bibitem[{\citenamefont{Vidmar et~al.}(2009)\citenamefont{Vidmar, Bon\v{c}a,
  Maekawa, and Tohyama}}]{bonca4}
\bibinfo{author}{\bibfnamefont{L.}~\bibnamefont{Vidmar}},
  \bibinfo{author}{\bibfnamefont{J.}~\bibnamefont{Bon\v{c}a}},
  \bibinfo{author}{\bibfnamefont{S.}~\bibnamefont{Maekawa}}, \bibnamefont{and}
  \bibinfo{author}{\bibfnamefont{T.}~\bibnamefont{Tohyama}},
  \bibinfo{journal}{Phys. Rev. Lett.} \textbf{\bibinfo{volume}{103}},
  \bibinfo{pages}{186401} (\bibinfo{year}{2009}).

\bibitem[{\citenamefont{Chernyshev et~al.}(1998)\citenamefont{Chernyshev,
  Leung, and Gooding}}]{cherny1}
\bibinfo{author}{\bibfnamefont{A.~L.} \bibnamefont{Chernyshev}},
  \bibinfo{author}{\bibfnamefont{P.~W.} \bibnamefont{Leung}}, \bibnamefont{and}
  \bibinfo{author}{\bibfnamefont{R.~J.} \bibnamefont{Gooding}},
  \bibinfo{journal}{Phys. Rev. B} \textbf{\bibinfo{volume}{58}},
  \bibinfo{pages}{13594} (\bibinfo{year}{1998}).

\bibitem[{\citenamefont{Mishchenko and Nagaosa}(2004)}]{nagaosa}
\bibinfo{author}{\bibfnamefont{A.~S.} \bibnamefont{Mishchenko}}
  \bibnamefont{and} \bibinfo{author}{\bibfnamefont{N.}~\bibnamefont{Nagaosa}},
  \bibinfo{journal}{Phys. Rev. Lett.} \textbf{\bibinfo{volume}{93}},
  \bibinfo{eid}{036402} (\bibinfo{year}{2004}).

\bibitem[{\citenamefont{Wr\'obel and Eder}(1998)}]{eder}
\bibinfo{author}{\bibfnamefont{P.}~\bibnamefont{Wr\'obel}} \bibnamefont{and}
  \bibinfo{author}{\bibfnamefont{R.}~\bibnamefont{Eder}},
  \bibinfo{journal}{Phys. Rev. B} \textbf{\bibinfo{volume}{58}},
  \bibinfo{pages}{15160} (\bibinfo{year}{1998}).

\bibitem[{\citenamefont{Riera and Dagotto}(1998)}]{riera2}
\bibinfo{author}{\bibfnamefont{J.}~\bibnamefont{Riera}} \bibnamefont{and}
  \bibinfo{author}{\bibfnamefont{E.}~\bibnamefont{Dagotto}},
  \bibinfo{journal}{Phys. Rev. B} \textbf{\bibinfo{volume}{57}},
  \bibinfo{pages}{8609} (\bibinfo{year}{1998}).

\bibitem[{\citenamefont{Sebastian et~al.}(2010)\citenamefont{Sebastian,
  Harrison, Altarawneh, Mielke, Liang, Bonn, Hardy, and Lonzarich}}]{sebastian}
\bibinfo{author}{\bibfnamefont{S.~E.} \bibnamefont{Sebastian}},
  \bibinfo{author}{\bibfnamefont{N.}~\bibnamefont{Harrison}},
  \bibinfo{author}{\bibfnamefont{M.~M.} \bibnamefont{Altarawneh}},
  \bibinfo{author}{\bibfnamefont{C.~H.} \bibnamefont{Mielke}},
  \bibinfo{author}{\bibfnamefont{R.}~\bibnamefont{Liang}},
  \bibinfo{author}{\bibfnamefont{D.~A.} \bibnamefont{Bonn}},
  \bibinfo{author}{\bibfnamefont{W.~N.} \bibnamefont{Hardy}}, \bibnamefont{and}
  \bibinfo{author}{\bibfnamefont{G.~G.} \bibnamefont{Lonzarich}},
  \bibinfo{journal}{PNAS} \textbf{\bibinfo{volume}{107}}, \bibinfo{pages}{6175}
  (\bibinfo{year}{2010}).

\end{thebibliography}

\end{document}